# Quantum Cryptography: an overview of Quantum Key Distribution

Davide Rusca

Vigo Quantum Communication Center, University of Vigo, Vigo E-36310, Spain.

Escuela de Ingenieria de Telecomunicacion, Department of Signal Theory and Communications, University of Vigo, Vigo E-36310, Spain.

AtlanTTic Research Center, University of Vigo, Vigo E-36310, Spain.

Nicolas Gisin

Group of Applied Physics, University of Geneva, 1211 Geneva 4, Switzerland

Constructor University, Bremen, Germany.

## Abstract
This chapter highlights the transformation of secure communications through the incorporation of quantum mechanics. Over the past four decades, this groundbreaking theory has quietly revolutionized private communication. The chapter provides a concise historical overview of this field's inception, tracking the development of its pioneering protocol, BB84. It delves deeply into the protocol's evolution, spotlighting its milestones and challenges. Furthermore, it offers a panoramic view of the entire quantum key distribution landscape, encompassing continuous variable protocols designed to harness existing telecom technologies and device-independent quantum key distribution protocols aimed at achieving secure key exchange with minimal reliance on the experimental setup.

## Introduction

Cryptography, one of the most enduring abstract technologies in human history, has roots dating back to the earliest need to safeguard secrets. Its historical traces can be discovered in the hieroglyphs of ancient Egypt (around 1800 B.C.) and the use of steganography among the ancient Greeks. Over time, it evolved into various cryptographic methods, with the substitution alphabet, exemplified by the Caesar cipher, becoming a well-known technique. To underscore the significance of cryptography, consider a widely recognized historical event: the encryption system used by Nazi Germany, known as the Enigma machine. This encryption system played a pivotal role in World War II. Its decryption by the brilliant mind of Alan Turing and his team played an indispensable role in the early and decisive victory of the Allied forces in the Second World War.

Until the late 1800s, cryptography was mostly about war and government secrets. But then, everything changed when new communication tech came along. First, there was the telephone, which ended up in people's homes, and then the super common cellphone, which slipped into everyone's pockets. This shift accelerated even further with the advent of the internet, allowing the world to achieve an unprecedented level of interconnectedness by the late twentieth century. With vast amounts of information traversing the globe at incredible speeds, privacy became a necessity not only for major corporations, governments, and banks but also for ordinary people. In today's world, virtually every communication, if adequately secured, relies on some form of cryptographic protection.

One of the most popular encryption schemes in the era of internet communication is the asymmetric key encryption method, with the RSA protocol being the most well-known and widely used (named after its authors, Ronald Rivest, Adi Shamir, and Leonard Adleman (Rivest et al., 1978)). This protocol operates on a straightforward principle. Instead of needing to establish a shared secret in advance between two parties looking to communicate securely, the sender generates two distinct binary strings: a public key and a private key. The public key is openly broadcasted to the receiver, making it susceptible to interception even by potential eavesdroppers. The receiver then employs the public key and the confidential message they wish to share, such as bank information, emails, or website content, and uses an algorithm to combine them before sending it back to the other party. While the public key and algorithm are known to potential adversaries, the essence of public key encryption lies in the fact that, for computationally constrained adversaries, decrypting the message becomes an incredibly challenging and time-consuming task. Conversely, it should be effortless for the party possessing the private key to decipher the secret message.

In the case of RSA encryption, the underlying problem it relies on is the famous factorization problem. For thousands of years, humans attempted to find an efficient algorithm to factorize large numbers, but they met with limited success. Unfortunately, the strength of this approach started to weaken in the 1990s with the introduction of an entirely new class of algorithms that surpassed what was previously thought to be achievable by a Turing machine.

The groundbreaking quantum algorithm developed by Peter Shor in 1994 (Shor, 1994) sent ripples of concern throughout the cryptographic community. Shor's work demonstrated that a quantum computer could efficiently factorize large prime numbers, effectively undermining what had long been considered a secure method for online communication, such as RSA encryption. You might wonder why this revelation didn't immediately lead to chaos, as it appeared to give adversaries the means to break widely used encryption. The reason for this apparent calm is that, at that time, and even today, we have not yet constructed a quantum computer with the requisite computational power to factorize numbers beyond what a high school student could manage with patience (Vandersypen et al., 2001). Building a practical quantum computer, capable of scaling to the level required for real-world applications, is an extraordinarily challenging task ("40 years of quantum computing," 2022). This journey to create such machines is a hotly contested race among various quantum technology approaches, each vying to become the first platform for a fully functional, general-purpose quantum computer. While the intricacies of how these machines work and the details of qubit creation and entanglement are subjects of significant depth, it's worth noting that most of the funding for quantum technologies is dedicated to this grand pursuit. This investment is driven not only by the potential for quantum computers to break encryption but also because they promise to revolutionize a wide array of applications that extend far beyond cryptography.

Amidst the excitement surrounding quantum computing, the rest of the cryptographic community is gripped by a sense of apprehension. There's a looming fear that a major player like IBM, Google, ETH, or others might announce the arrival of a quantum computer capable of breaking existing encryption. This prospect raises concerns about a potential information apocalypse. In response, the cryptographic community is pursuing two main approaches, each with its own set of advantages and drawbacks. The first approach, known as Post-Quantum Cryptography (PQC) or Quantum-Resilient Cryptography (Bernstein and Lange, 2017), involves sticking with the public key encryption model but using different types of algorithms. These algorithms are chosen not only for their classical computational complexity but also because there are no known quantum algorithms capable of efficiently solving them. The main assumption here is that, aside from a computationally bounded adversary, PQC relies on the belief that no quantum algorithm, more efficient than classical ones, can be devised to break these encryption methods. On the other hand, Quantum Cryptography takes a fundamentally different approach. It leverages the principles of quantum mechanics to attempt secure information exchange. The most well-known quantum cryptographic scheme is Quantum Key Distribution (QKD), often paired with classical symmetric key encryption methods to achieve information theoretical security in encrypting and transmitting a message.

## Quantum Key Distribution

Quantum Key Distribution (QKD) addresses a problem that classical information theory struggled to resolve. Claude Shannon, in a seminal work from 1949 (Shannon, 1949), demonstrated that a symmetric scheme known today as the "one-time pad" is information theoretically secure. In simpler terms, this means that if an adversary intercepts the ciphertext (the encrypted message), they gain no information about the plaintext (the original message). To them, the ciphertext appears indistinguishable from a random string of text. The concept behind the one-time pad is elegantly simple. When encoding a message as a binary string of length 'n,' a secret and random bit string of the same length is generated. A basic Boolean operation called XOR (exclusive or) is then applied to each corresponding pair of bits in the two sequences, creating the ciphertext. It's clear that if the bits in the secret key are genuinely random and kept secret, XOR-ing them with any other sequence will result in a ciphertext that appears random to anyone without access to the key. However, a challenge arises when the other party needs to decipher the message. They must possess the same key used for encryption. Herein lies the problem: distributing a key if the message is non-trivial. As the attentive reader may have noticed, with classical information theory, it's impossible to remotely distribute such keys. This implies that two parties opting for the one-time pad would need to exchange enough keys before moving to remote locations. The catch is that once these keys are exhausted (as the name "one-time pad" suggests, each key can be used only once), they would need to meet in person again, which is often impractical.

Quantum mechanics provides an ingenious solution to this challenge, allowing for the continuous exchange of keys between two remote parties. Interestingly, the foundation for this field was laid a decade before the introduction of Shor's algorithm. In this chapter, we'll embark on a journey through the birth and evolution of quantum key distribution (QKD). Our approach will primarily be chronological. We'll begin by introducing the original QKD protocol, presented in 1984 by Bennett and Brassard. From there, we'll delve into the subsequent changes and adaptations that emerged because of the real-world challenges faced when implementing quantum key distribution. It's essential to note that QKD, and quantum cryptography, are in a state of continuous improvement. Attempting to encompass all developments in a single chapter would be a formidable task and many reviews of the field do that extensively (Diamanti and Leverrier, 2015; Gisin et al., 2002; Pirandola et al., 2020; Xu et al., 2020). Instead, we aim to present the key milestones in this field and provide readers with ample references to satisfy any lingering curiosity.

## The start of QKD: the BB84 protocol

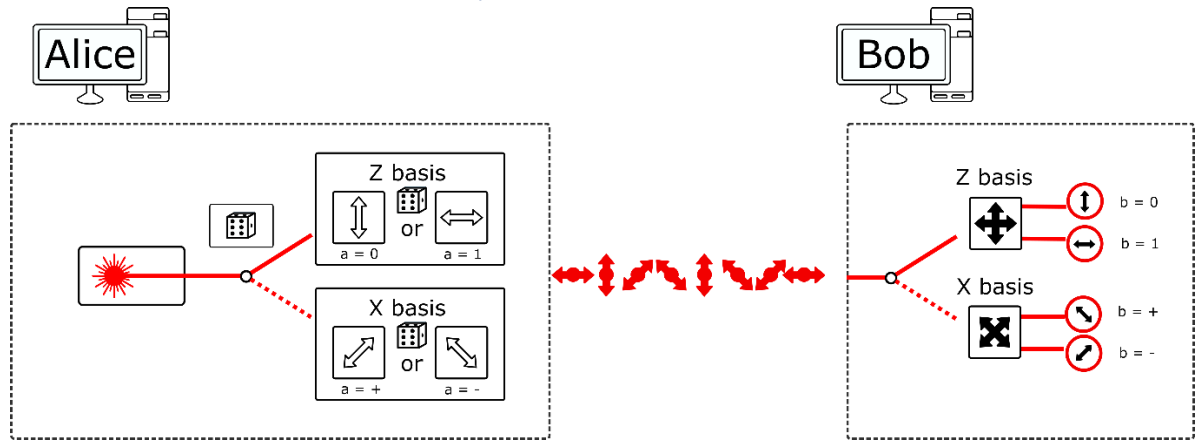

Figure 1: Scheme of the BB84 protocol. Alice choses at random the basis, and once the basis is chosen selects at random one of the two orthogonal states in the basis and sends it to Bob. Bob choses at random one of the two basis to perform the measurement.

In 1984, Charles Bennett and Gilles Brassard introduced one of the cornerstones of the second quantum revolution—a protocol for quantum key distribution now known as BB84, named in their honor (Bennett and Brassard, 1984). Their work built upon the foundations laid by Stephen Wiesner in the 1970s, particularly his exploration of quantum conjugate encoding. Wiesner's early work introduced the concept of encoding information in two conjugate bases within the qubit space (Wiesner, 1983). Bennett and Brassard found this concept intriguing, but instead of utilizing qubits solely for information storage, they delved into their potential for secure information transmission.

The essence of BB84 lies in using qubits to transmit a secret key (see Figure 1). They make use of two conjugate bases within the Hilbert space of dimension 2, commonly referred to as the qubit space. For simplicity, we will employ two bases throughout this chapter: the Z basis, represented by $|0\rangle$ and $|1\rangle$, and the X basis, represented by $|+\rangle = \frac{|0\rangle+|1\rangle}{\sqrt{2}}$ and $|-\rangle = \frac{|0\rangle-|1\rangle}{\sqrt{2}}$. Here's how the BB84 protocol works: Alice randomly selects a basis and a state within that basis. She then sends the qubit to Bob through a completely untrusted channel, where an adversary (typically referred to as Eve) can employ any quantum mechanical operation to manipulate the states passing through. Bob, upon receiving the qubit, chooses a basis for measurement. If Alice and Bob both select the same basis, and there's no interference or loss in the channel, they will obtain the same measurement result, as the state sent corresponds to the state received. However, if their chosen bases do not align, quantum mechanics dictates that the measurement outcomes are completely uncorrelated. For instance, if Alice sends the state $|0\rangle$ in the Z basis and Bob measures in the X basis, the Born rule tells us that Bob will obtain either $|+\rangle$ or $|-\rangle$ with equal probability of $1/2$. Consequently, events in which the two honest parties have chosen different bases are discarded in the BB84 protocol.

In the ideal scenario, it's straightforward to see that we can share a random key (random because Alice's bit choices are random, and it's random whether Bob chooses the same basis as her) correct up to the noise of the channel (with losses being discarded as inconclusive events, similar to different basis choices). However, ensuring that Eve cannot extract any information from the channel is not a trivial task. It took the quantum cryptography community nearly two decades to reach a definitive solution to this problem, at least in the ideal scenario (Shor and Preskill, 2000) (we'll return to what we mean by "ideal" later). Intuitively, the security of BB84 relies on the unique properties of quantum mechanics. Even the simplest attack, where Eve intercepts and resends qubits, leaves a distinct mark on the transmission. This is because Eve does not know which basis Alice chose, and if the probabilities for Z or X are equal at ½, then that's also the probability that Eve guesses correctly. When Eve guesses right, she becomes completely correlated with Alice, producing no errors. However, if she fails, she will resend a state in the wrong basis, causing an error of 50%. Consequently, regardless of the channels, Alice and Bob will estimate an average error rate of 25% due to Eve's interference.

But what if the adversary attempts to copy the qubits and store them until the public disclosure of the basis, allowing Eve to measure them in the correct one? This approach, unfortunately, is not feasible without interfering with the transmission. Unlike classical bits, Eve cannot copy the stream of qubits that Alice sends without introducing errors. This is because of the no-cloning theorem, which states that it's impossible to create an exact copy of an arbitrary unknown quantum state from a non-orthogonal set of states (if they were all orthogonal, they could be treated as classical bits of information, distinguishable perfectly by a single basis).

Indeed, these two examples serve as compelling arguments for why the BB84 protocol should work, but they are not definitive proofs of its security. To address more intricate questions, such as what happens if Eve selectively attacks only certain rounds, the quantum cryptography community has undertaken extensive research and analysis.

It's important to deter the reader from advocating a zero-error policy, which would allow for key exchange only in the absence of any errors during communication. This approach is impractical, as the transmission of qubits occurs at a physical layer of the communication system, which is inherently susceptible to noise and imperfections. Consequently, perfect error-free transmission is rarely achievable, making it necessary to develop protocols that can still securely exchange keys even in the presence of some errors. This recognition of the practical challenges and imperfections inherent to real-world systems is crucial in the field of quantum key distribution.

Regarded as the firsts complete and feasible security proofs, the works of (Mayers, 2001; Shor and Preskill, 2000) gives the dependence of the secret key rate, the amount of key that can be distilled and the different error rates:

*Equation 1*

$$r = R(1 - H(Q_z) - H(Q_x))$$

where R is the detection rate of our qubits, $Q_z$ and $Q_x$ are the error rates respectively on the Z and X basis, and $H(Q) = -q \log_2 q - (1-q) \log_2(1-q)$ is the binary entropy. From this simple relation we can understand two important concepts. First, the more loss we have in the channel the lower the key exchange rate will be since R will decrease. Moreover, we see that if the error rate increases, the key rate decreases, meaning that since quantum mechanics grants us the property that if Eve tries to measure the qubits the error rate will increase. Any attempt to tamper with the communication will result in a reduction of the secret key rate, until no key exchange can be performed. In other words, if Eve tries to attack the channel, we will be able to detect it and act accordingly.

Recently a modern way to look at the security of QKD relies on the definitions of composability and $\varepsilon$ security. Intuitively, the idea is to have a protocol that is indistinguishable up to an arbitrary probability $\varepsilon$ to an ideal key distribution scheme (for more info on the matter refer to (Renner, 2006)).

## First QKD implementations: form single photons to coherent states

To turn this abstract scheme into an experiment, researchers in the field of quantum key distribution turned to a brilliant solution pioneered by Stephen Wiesner in the 1970s: the use of photon polarization. Two orthogonal polarizations of light form a Hilbert space of dimension 2, essentially creating a qubit space, when considering only one photon at a time. The superposition of these polarizations naturally establishes the other basis required for the BB84 protocol. Moreover, light serves as an excellent medium for transmitting information, whether through optical fibers (as in most classical communication) or in free space. As it turns out, polarization is just one of the possible ways to achieve this. What is sufficient is to find two orthogonal modes of light, in which a photon can be encoded, separately and in their superposition.

However, there's a challenge to overcome, that of producing single photons. While generating light is a straightforward task that humanity has mastered since the discovery of fire, creating single photons is far more complex. The condition for BB84 to function correctly is the absence of multiphoton states in the prepared states. If multiphoton states are present, an adversary could intercept the surplus photons and obtain a copy of the original state, thereby violating the security measures mentioned earlier.

In the 1980s and 1990s, single photon sources were not readily available. Even today, the state-of-the-art single photon sources are not perfect, as they produce states that are close to single photons (Eisaman et al., 2011; Lounis and Orrit, 2005; Meyer-Scott et al., 2020). These sources also perform less efficiently compared to other light sources, such as lasers, which can be modulated to generate temporal states at repetition rates orders of magnitude higher than those of single photons. Nevertheless, the use of single photons in QKD experiments is a crucial element in ensuring the security of quantum key distribution systems.

The first implementation of BB84 was carried out using dim LED instead of a single photon source (Bennett and Brassard, 1989). Shortly after lasers (coherent states) were proposed for the same purpose (Huttner et al., 1995), quickly followed by many implementations (Bourennane et al., 1999; Chiangga et al., 1999; Hughes et al., n.d.; Ribordy et al., n.d.; Townsend, 1998). The hope was that a security proof for this modified protocol would yield performances comparable to those of the protocol using single photons.

Unfortunately, this was not the case. A first attack would was found in the early 2000s (Brassard et al., 2000; Lütkenhaus, 2000) in which Eve could take advantage of the multiphoton nature of the state sent by Alice by splitting the power sent from Alice to Bob with a beam splitter and fill the distance between her and the trusted receiver with a lossless channel. Eve in fact is limited only by the laws of physics in their attacks.

Whilst this first attack seemed not too effective, another one was found just after. In 2002, (Lütkenhaus and Jahma, 2002) devised a new attack on the coherent BB84 protocol (that was already mentioned in (Huttner et al., 1995)) that proved to be significantly more effective than previous attacks. Their approach hinged on exploiting the fact that a coherent state is a superposition of number states, and the probability of measuring a certain number of photons in a coherent state follows a Poisson distribution.

*Equation 2*

$$|\alpha\rangle = e^{-\frac{|\alpha|^2}{2}} \sum_{n=0}^{\infty} \frac{\alpha^n}{\sqrt{n!}} |n\rangle$$

and

*Equation 3*

$$p_n = |\langle n|\alpha\rangle|^2 = e^{-|\alpha|^2} \frac{|\alpha|^{2n}}{n!}$$

where $\alpha$ is the complex number characterizing the coherent state and $p_n$ is the probability of detecting n photons given that the state $|\alpha\rangle$ was sent. As an example, for a coherent state with average photon number $|\alpha|^2 = 0.5$ the probability to detect a multiphoton state is

*Equation 4*

$$p_{multi} = 1 - p_0 - p_1 = 1 - e^{-|\alpha|^2} - e^{-|\alpha|^2}|\alpha|^2 = 1 - e^{-0.5} - e^{-0.5}0.5 \cong 0.09$$

which means that almost 10% of the time we would detect a multiphoton state.

The essence of this attack was to make use of these multi-photon events. Instead of merely splitting the incoming light using a basic beam splitter, the adversary, just outside Alice's lab, performed a non-disruptive measurement of the number of photons in a pulse. Importantly, the number of photons measurement is orthogonal to any canonical encoding basis chosen by the honest parties (Calsamiglia et al., 2001).

The adversary's strategy then entails measuring the photon number of each state passing through the channel and halting every single photon they can (ensuring that the detection matches the actual channel loss). They also "tag" every multi-photon state. In this context, tagging means storing a portion of the multi-photon state in memory (until classical communication is complete) and letting the remainder pass through a lossless channel to Bob. This means that any state "tagged" by the adversary is going to be known by them after the basis are disclosed since they are going to measure their stored part in the same bases as Alice and Bob.

While this tactic poses a significant threat to the protocol's security, it does not completely break it. The maximum achievable distance is limited by the number of multi-photons sent. Therefore, for longer transmission distances, the average photon number in the coherent state must be reduced. However, if the intensity of the state is reduced to extend the transmission range, it implies that the secret key rate

(SKR) will decrease quadratically with respect to the losses (for more details, refer to the original paper (Lütkenhaus and Jahma, 2002)).

## Protocols resilient to the PNS attack

In the following years many attempts were made to overcome this problem.

One notable protocol developed in the quest for improved security was SARG04, introduced by Scarani, Acin, Ribordy, and Gisin in 2004 (Scarani et al., 2004). SARG04 utilized the same experimental setup as the BB84 protocol, but with a different classical disclosure method between Alice and Bob.

In the SARG04 protocol, Alice grouped the possible states she would send into pairs: $|0\rangle$ and $|+\rangle$, and $|1\rangle$ and $|-\rangle$. She would disclose which set she sent to Bob only after the measurement phase was completed. Bob, in turn, would choose to measure either in the Z or X basis. He would inform Alice that a measurement was successful only if the measurement result did not belong to the set disclosed by Alice. For example, if Bob measured in the X basis and obtained $|-\rangle$ as a result, and Alice disclosed that she sent either $|0\rangle$ or $|+\rangle$, Bob would infer that Alice had sent $|0\rangle$, as no other possibility would yield the observed result in the absence of errors.

The intention behind this protocol was to enhance its resilience against photon number splitting (PNS) attacks, particularly when applied to a coherent implementation, as there was no immediate disclosure of the basis by Alice. However, as demonstrated in the security analysis by (Fung et al., 2006), the protocol's behavior against general attacks did not exhibit a significant advantage over the BB84 protocol.

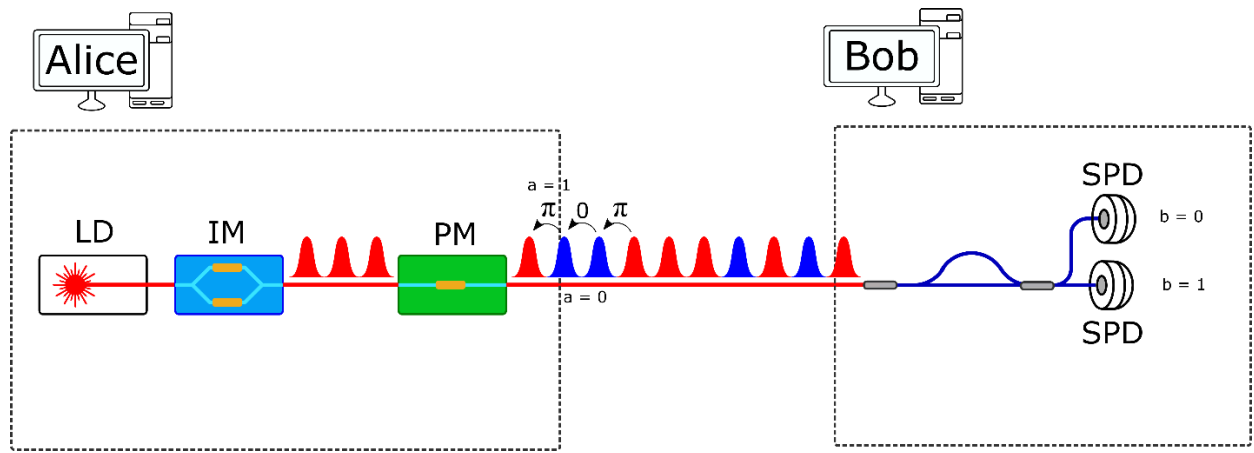

Figure 2 Experimental scheme for differential phase shift protocol.

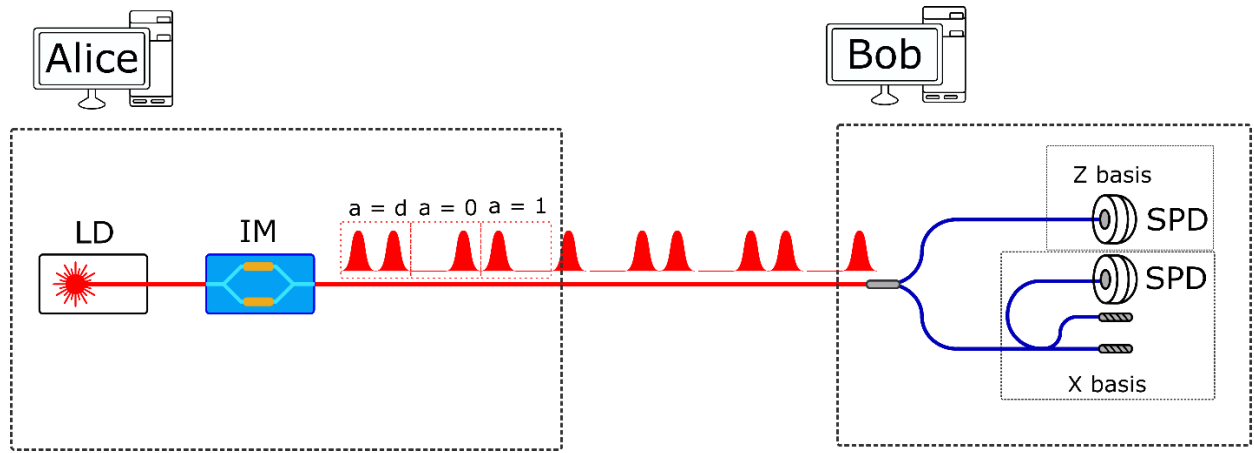

*Figure 3 Experimental scheme for the coherent one way protocol.*

Another avenue explored in the pursuit of more secure quantum key distribution protocols is the class of protocols known as distributed phase protocols. Over the years, various protocols have emerged in this category, including Distributed Phase Shifting (DPS) (Inoue et al., 2002), a variation known as the Round Robin (RR) (Sasaki et al., 2014), and the Coherent One Way (COW) protocol (Stucki et al., 2005). While these protocols employ different methods to encode and transmit keys, what unites them is the principle of distributed coherence between different rounds of the protocol.

In the DPS protocol (see Figure 2), a series of coherent pulses is generated, and the phase of each pulse is modulated between 0 and π. On Bob's side, an interference measurement is conducted for each pulse. With information encoded between each pulse, if a photon number-resolving measurement is performed, the coherence between two different rounds is disrupted, allowing the honest parties to detect any eavesdropping.

The Coherent One Way protocol (see Figure 3), on the other hand, modulates the intensity of the pulses instead of the phase. By checking the coherence between two consecutive qubits, it serves a similar purpose to the DPS protocol. However, the added measurement between each round complicates the task of developing a security proof. Many of the conventional techniques used for security proofs are not easily applicable in these protocols, which has left them with incomplete proofs and potential vulnerabilities against new attacks for a significant period.

While we have made progress in developing the necessary techniques for the security of these protocols today, recent results suggest that their secret key rate (SKR) scaling is more like the Coherent BB84 protocol than to the original single photon version. This demonstrates the ongoing challenge of balancing security and practicality in the development of quantum key distribution protocols.

## Decoy state BB84

The community's approach to enhancing the security of the BB84 protocol led to the development of what is known today as the decoy-state BB84 protocol. This protocol was formulated in the early 2000s and remains one of the most widely used protocols in both academic and commercial contexts (Hayashi and Nakayama, 2014; Hwang, 2003; Lim et al., 2014; Lo et al., 2005; Ma et al., 2005; Wang, 2005).

The core concept of the decoy-state BB84 protocol involves slight modifications to the coherent BB84 protocol, aimed at detecting potential photon number splitting (PNS) attacks. Instead of using a fixed average photon number for her states, Alice in the decoy-state BB84 protocol has the choice between different values. She randomizes the phase between each pulse, allowing each state to be expressed not as a coherent superposition of number states but as a classical mixture of them. This means that the state preparation corresponds to Alice choosing a Poissonian distribution with an average of $\mu_i$ and sending a number state according to this distribution. This corresponds to the state with a density matrix as follows:

$$\rho_\mu = e^{-\mu} \sum_{n=0}^{\infty} \frac{\mu^n}{n!} |n\rangle\langle n|$$

where for simplicity we wrote the average photon number as $\mu = |\alpha|^2$.

Now, if Eve attempts a PNS attack, her challenge is that indiscriminately stopping single photons will alter the detection probabilities of the two distributions in distinct ways, making it possible for the honest parties to detect her interference.

The crucial point in the protocol is that, due to the classical mixture of number states, the total number of detections at Bob can be decomposed into the sum of the probability to send a particular photon number state and the number of detections associated with that state. After Alice discloses to Bob which intensity she chose, the protocol ends with a system of linear equations. These equations have as many variables as the number of detections depending on each separate state. However, the only significant number is the count of single photon detections, as this reveals how many single photons were sent by Alice, traversed the channel, and arrived at Bob.

The security proof for the decoy-state BB84 protocol applies the same principles as for single photon states. In a reference (Lim et al., 2014), it was demonstrated that using three intensity values is optimal for estimating a lower bound on the number of single photon detections. This provides the protocol with a robust security foundation, ensuring its effectiveness against various potential attacks.

This protocol has nowadays been implemented many times in many different configurations, achieving both long communication distance in fiber (Boaron et al., 2018) and in free space/satellite (Liao et al., 2017) and fast key generation rate of more than tens of Mbits/s (Grünenfelder et al., 2023; Li et al., 2023; Yuan et al., 2018)

## Continuous variable QKD

In the previous sections, we discussed various quantum key distribution (QKD) protocols, with a particular focus on BB84 and its derivative protocols. These protocols fall under the category of discrete variable protocols (DV-QKD), as they rely on discrete observables for encoding and detecting information.

While BB84 could theoretically be implemented with any Hilbert space of dimension two, it became clear early on that light, especially its orthogonal modes like polarizations, would be the most practical means of implementation. However, in the late 1990s and early 2000s, researchers began exploring alternative ways to utilize light for quantum communication. This exploration started with the works of (Ralph, 2000, 1999) and centered on the field observables of light, based on the quadrature operators $\hat{q}$ and $\hat{p}$, where:

$$\hat{q} := \hat{a} + \hat{a}^\dagger \quad and \quad \hat{p} := i(\hat{a} - \hat{a}^\dagger)$$

Coherent states emerged as optimal candidates for encoding information using quadrature observables. Coherent states exhibit evenly distributed uncertainties between the two quadratures, X and P, while simultaneously minimizing the uncertainty relationship between these observables. Furthermore, for measuring the quadratures, simple shot noise-limited homodyne and heterodyne measurements are already highly effective. These measurements rely on balanced linear detectors, as opposed to more complex single photon detectors commonly used in discrete variable QKD. This shift towards continuous variable QKD opened new possibilities for quantum communication using light.

The advantage of continuous variable quantum key distribution (CV-QKD) is evident, as it allows the use of readily available coherent sources like lasers without the need for additional steps in the protocol, such as those required for the decoy state method in discrete variable QKD (DV-QKD). Furthermore, the detection schemes used in optical homodyne measurements can be borrowed from classical telecommunications, where quadrature amplitude and phase modulation are standard practices. This approach essentially leverages existing technology from classical communications to implement quantum key distribution, simplifying the implementation aspect.

However, while CV-QKD simplifies implementation, it introduces a more complex analysis in terms of security. When dealing with quadrature and field operators, the Hilbert spaces involved are infinite-dimensional, which presents a significant challenge in developing definitive security proofs for the most popular CV-QKD protocols. Moreover, the equivalence of vacuum and noise for homodyne/heterodyne detection corresponds to an increasing quantum bit error rate with respect to the loss of the channel, i.e. every round a conclusive event is required (each lost photon becomes an error), which has as a consequence that a higher loss increases the error rate until almost 50%. This requires heavy error correcting codes that can support high error rates, whilst for DV QKD the error rate stays more or less constant with respect to loss.

Like DV-QKD, where many protocols were proposed over the years but only a couple remain of interest today, the same holds true for CV-QKD. The fundamental formula remains largely consistent across most of the protocols currently in use. In CV-QKD, Alice generates and sends coherent states to Bob following a specific distribution, where the complex values α = q + ip are selected based on this distribution. Information is encoded in the quadratures of these states and transmitted to Bob. Bob can perform simultaneous measurements of both quadratures using heterodyne detection or alternate measurements of the two quadratures using homodyne detection. By analyzing the variance of the distribution of x and p both before (generated by Alice) and after (measured by Bob) transmission, it is possible to estimate the amount of information leaked to Eve. With this information, error correction and privacy amplification techniques can be implemented to achieve composable ε-security.

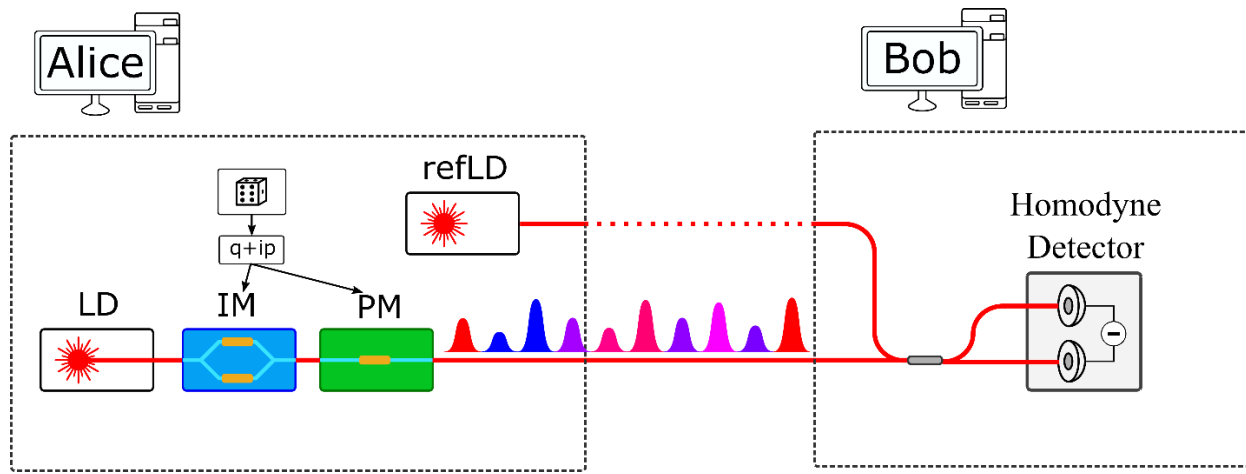

*Figure 4 Scheme of a typical CV QKD experiment, Alice choses a complex number q + ip following a certain distribution (gaussian modulation, discrete modulation) and prepares a coherent state by using an intensity modulator and a phase modulator. A shared laser reference between Alice and Bob it is either sent through a physical channel, or it is synchronized remotely between the two honest parties.*

In the two main CV-QKD protocols in use today, modulation of the coherent states in the phase space can be done continuously, by modulating them according to a Gaussian distribution with a mean of zero, or discretely, using typical quadrature phase shift modulation (Guo et al., 2018; Silberhorn et al., 2002). These methods help to ensure the security and reliability of CV-QKD communication continuously improving the performance and state of the art of this technique (Huang et al., 2015; Kumar et al., 2015; Zhang et al., 2020).

## Side channels and Hacking in Quantum Cryptography

While the first part of this chapter might lead the reader to believe that quantum key distribution (QKD) is an ultimate, unbreakable form of cryptography, the reality is that every deviation from the ideal protocol can introduce potential vulnerabilities. This is true for QKD, as well as for any cryptographic system. Hackers and malicious parties often focus on finding weaknesses in the implementation of protocols rather than attempting to directly crack well-established, theoretically secure ones.

For QKD, the challenge of implementation security is one of the most intriguing aspects, both from an academic perspective (addressing device imperfections and potential side channels within security proofs) and from a practical standpoint (developing countermeasures to ensure the security of existing protocols).

To provide the reader with insight into the nature of these attacks, let's focus on two of the most notorious types of attacks: those targeting the sender and those targeting the receiver.

Attacks on the sender are typically aimed at eavesdropping on the settings chosen by Alice without tampering with the quantum channel itself. One of the most common attacks of this kind is the Trojan horse attack (Gisin et al., 2006; Jain et al., 2015, 2014; Vakhitov et al., 2001), in which the attacker sends a pulse of light into Alice's device at a different wavelength than that of the quantum channel, to avoid detection. The reflected portion of the injected light becomes modulated in the same way as the light used in the QKD protocol but at a different wavelength, making it difficult for the honest parties to detect the attack. Countermeasures for this type of attack often involve implementing an optical isolator, which significantly increases the loss for counterpropagating fields inside the transmitter. By doing so, the power injected by the attacker cannot be excessively high, as very high power levels would damage standard optical fibers and trigger an aborted condition between Alice and Bob. This countermeasure aims to minimize the amount of light and information that a malicious party can steal. However, as is the case with any attack and countermeasure, determined attackers can seek ways to bypass these countermeasures. For instance, in a particular study (Makarov et al., 2016) the authors attempted to damage the optical isolator, allowing for a higher amount of reflected light to escape from the sender.

The battle between attacks and countermeasures in QKD serves as a testament to the ongoing quest to ensure the security of quantum communication protocols in practical, real-world implementations.

The receiver has long been recognized as the most vulnerable component of a quantum key distribution (QKD) system. This vulnerability becomes apparent when considering countermeasures against light injection attacks. Unlike the sender, the receiver cannot rely on optical isolators to mitigate potential light injection attacks because such countermeasures would lead to significant losses in the quantum channel, making communication between the legitimate parties impossible. As a result, the receiver is susceptible to various forms of light injection attacks, particularly when it comes to the detectors (Bugge et al., 2014; Gerhardt et al., 2011; Lydersen et al., 2010; Makarov, 2009).

In QKD security proofs, detectors are often treated as threshold detectors. In principle, they should ideally detect an input state with some level of noise and imperfections. However, this ideal behavior is applicable primarily to dim light input states. When the input state contains considerably more photons than anticipated, the behavior of single-photon detectors (SPDs), such as semiconductor SPDs can change (Eisaman et al., 2011). Similar principles apply to both avalanche photodiodes (APDs), and superconducting nanowire single-photon detectors (SNSPDs).

For example, in semiconductor SPDs, when a photon is detected, it triggers an avalanche of charges, resulting in a readable electrical current. This avalanche is then quenched, either passively or actively, allowing the voltage in the diode to recover above the breakdown level. However, when continuously exposed to a high number of photons (at a macroscopic level), the voltage never reaches a level where a full avalanche is initiated. In this regime, the device behaves differently: its output signal is no longer binary but is proportional to the incoming light intensity.

This alteration of detector behavior allows for a so-called "intercept and resend" attack. In this attack, Eve controls both the basis selection and state detection at the receiver. She intercepts a photon, selects a basis, and detects a state. Then, she manipulates Bob's detection process, forcing him to obtain either a vacuum event or the same detected state she observed. Importantly, this attack doesn't cause interference in the communication itself, but it completely correlates Eve's information with that of Alice and Bob.

Countermeasures against this attack typically involve monitoring the power entering Bob's device, looking for unusual patterns or excessive light, and potentially setting thresholds to detect and mitigate such attacks in real-time. These countermeasures aim to ensure that Bob's detectors operate in the ideal regime, where they behave as threshold detectors and are not susceptible to being manipulated by an adversary's injected light. However, implementing these countermeasures effectively requires vigilance and continuous monitoring to safeguard the security of the QKD system.

## Measurement Device Independent QKD

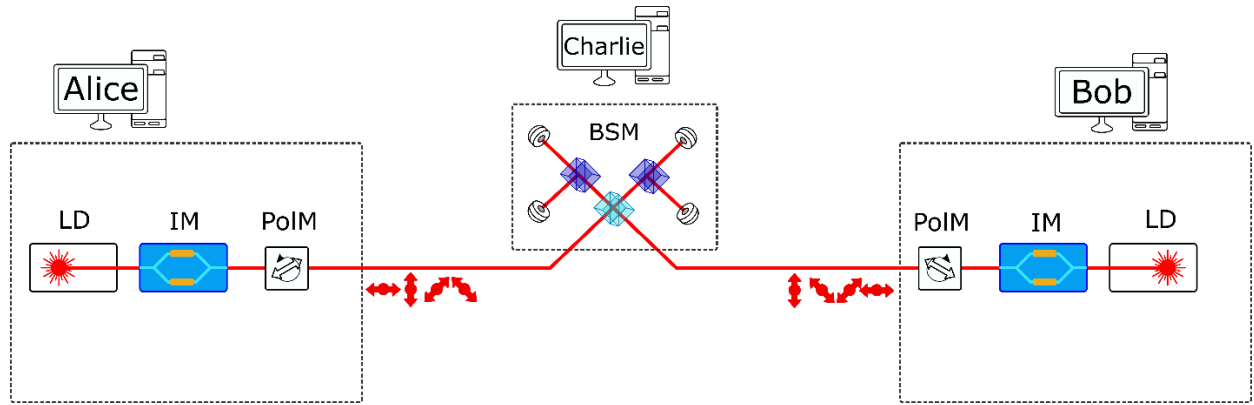

*Figure 5 Experimental implementation of a measurement device independent QKD.*

Measurement Device Independent Quantum Key Distribution (MDI-QKD) is an innovative quantum key distribution protocol that significantly enhances security by removing the need to characterize the detectors used in the system. This protocol was first presented in 2012 and represents a promising approach to secure quantum communication (Braunstein and Pirandola, 2012; Lo et al., 2012).

In MDI-QKD, both Alice and Bob, who are typically the sender and receiver in other QKD schemes like BB84, act as senders in the protocol. They could choose one of the two polarization bases used in BB84 and prepare one of the four states typically associated with BB84. These states are then sent to a central station, which can be regarded as uncharacterized, or even as a potential adversary (Eve). The central node is expected to perform a Bell state measurement using the inputs it receives from Alice and Bob.

The key feature of MDI-QKD is that knowing the result of the Bell state measurement at the central station does not provide Eve with sufficient information to compromise the security of the communication. However, both Alice and Bob can independently reconstruct the states sent by the other party, allowing them to correlate their measurement results and generate a shared secret key, just as in more conventional BB84-based schemes.

One important advantage of MDI-QKD is its immunity to general attacks, as the detectors are left in complete control of the adversary (the central node, which could be Eve). Consequently, any side-channel attack related to detector manipulation does not impact the security of the protocol. This enhances overall security by eliminating vulnerabilities related to the detectors, which are typically the most vulnerable parts of a QKD system.

It is important to note that while MDI-QKD offers significant security benefits, it does come with a more complex implementation. Precise synchronization between Alice and Bob is required, and laser wavelength locking is necessary for the protocol to work effectively. Despite these implementation challenges, the protocol's security advantages make it an appealing choice for secure quantum communication. Moreover, since its presentation (Ferreira Da Silva et al., 2013; Liu et al., 2013; Rubenok et al., 2013) MDI QKD, have been implemented in many successful experiments, reaching distances comparable to prepare and measure QKD (Wei et al., 2020; Yin et al., 2016).

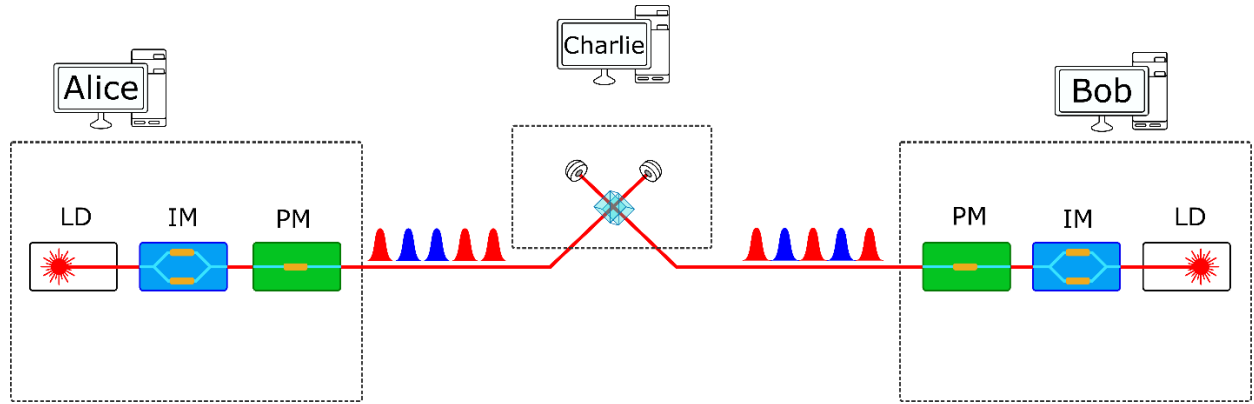

*Figure 6 Experimental implementation of a twin field quantum key distribution protocol.*

Another milestone was achieved in MDI-QKD with the introduction of Twin-field Quantum Key Distribution (TF-QKD). TF-QKD is a remarkable protocol introduced in 2018 (Lucamarini et al., 2018) that overcomes the linear scaling with loss limitation observed in previous repeaterless schemes (Pirandola et al., 2017), like the Measurement Device Independent QKD (MDI-QKD). It achieves this feat by using single photon interference rather than two-photon interference, which opens the door to a new era of secure long-distance quantum communication.

In the context of Twin-field QKD, imagine an experiment akin to a large interferometer, with Alice and Bob modulating each arm of this setup. When a single photon is injected into the interferometer, the collective action of Alice and Bob will determine whether it arrives at the right or left detector. Like MDI-QKD, revealing the measurement outcome alone doesn't provide enough information to deduce the precise modulation settings chosen by Alice and Bob. However, if one side's modulation setting is known, the other party's setting can be inferred directly.

The original proposal demonstrated that this effect could be achieved using two lasers phase-locked to each other, combined with the same central detection scheme. By modulating these laser states with 0 and π phase, Twin-field QKD offered security comparable to MDI-QKD but with a different scaling of the rate concerning distance. This protocol exhibits a square root scaling with loss, potentially extending the maximum achievable communication distance while utilizing the same detectors, modulators, and repetition rates. The price for this improvement in performance is an increase in experimental complexity. Achieving precise single photon interference necessitates not only frequency locking of the two lasers at Alice and Bob but also phase locking, which demands considerable technological effort. Moreover, stability in phase locking must be maintained over the connecting channels between Alice and Bob, which, for long-distance quantum communication spanning hundreds of kilometers, is non-trivial and introduces significant technical challenges.

Since its initial introduction, various proposals and practical solutions have been developed (Fang et al., 2020; Minder et al., 2019; Wang et al., 2019), achieving impressive results and record-breaking distances for fiber-based quantum communication. Innovations in laser phase locking techniques and improved methods for locking the phase of the equivalent interferometer, sometimes through post-processing corrections instead of locked-loop control, have allowed for long-distance secure quantum

communication(Pittaluga et al., 2021; Wang et al., 2022). Notably, these advancements have led to recent achievements in long-distance quantum communication, including the demonstration of over 1000 kilometers of real fiber communication (Liu et al., 2023), a feat previously deemed impossible. Twin-field QKD has thus ushered in a new era of secure and long-distance quantum key distribution.

## Entanglement Based QKD and DI QKD

Entanglement-based Quantum Key Distribution (QKD) has a rich history, dating back to the early 1990s when Ekert and Bennett, Brassard, and others presented pioneering schemes that harnessed the power of entanglement to enhance the security of quantum communication.

The BBM92 scheme, introduced in 1992 (Bennett et al., 1992), demonstrated that entanglement can be utilized to achieve the same level of security as the prepare-and-measure counterpart of BB84. It highlighted the use of source substitution and equivalence models, which are often employed in QKD to facilitate the security analysis of prepare-and-measure protocols. These techniques enable security proofs for protocols that would otherwise be challenging to analyze. The E91 protocol (Ekert, 1991), on the other hand, introduced an intriguing concept that would later contribute to a new level of implementation security in QKD.

In the E91 protocol, fully entangled states are distributed to both Alice and Bob. Occasionally, the two parties measure these states in the same computational basis, and their results would, by chance, be perfectly correlated in a noiseless scenario. The remaining measurements are performed to violate a CHSH inequality, the same inequality used in Bell nonlocality tests. Instead of testing the existence of quantum nonlocal behavior, the protocol assumes the validity of quantum mechanics and verifies whether any tampering has occurred with the test being conducted. Since entanglement is monogamous, a full violation of the CHSH inequality serves as evidence that the state supplied to Alice and Bob was genuinely an entangled state, proving the security of the protocol.

For many years, these results were deemed sufficient for academia. However, it was discovered that by rigorously closing all possible loopholes in Bell tests (locality loopholes, free will loopholes, and, most crucially, detection loopholes), the self-testing of entanglement and the security of the protocol could be achieved with minimal assumptions about the devices (Acín et al., 2007; Barrett et al., 2005; Pironio et al., 2009). In this device-independent quantum key distribution (DI-QKD), the behavior of the devices can be almost arbitrary, with certain limitations such as preventing the adversary from directly accessing the measurement outcomes, as such access would enable key theft. Nonetheless, the behavior of the devices can remain largely unchanged, allowing for the secure implementation of QKD. This revolutionary technique has opened new horizons in quantum key distribution by providing a higher level of device independence and security.

Due to the high complexity required by the experiment to achieve a successful exchange in the device-independent scenario, (a loophole free bell test must be performed) for a long time this protocol was considered impossible to implement. However, recently two groups managed to implement it experimentally (Nadlinger et al., 2022; Zhang et al., 2022).

## Conclusion

In this chapter, we have explored the fascinating world of quantum communication, with a focus on quantum key distribution (QKD). This field represents one of the most exciting areas in quantum technology today, where the promises of a more secure future are being realized through the availability of commercial quantum devices that operate at practical performance levels.

Quantum cryptography, a subset of quantum communication, marks the initial step in the second quantum revolution. It accomplishes feats that were once thought impossible in classical information theory. Remote secret key distribution, a fundamental application of QKD, is just one of the groundbreaking achievements in this field. Moreover, the development of QKD protocols like Measurement-Device-Independent (MDI) and Device-Independent (DI) schemes has expanded the horizons of device independence and practical security in quantum communication.

While quantum cryptography is transitioning toward industrial implementation and gaining traction in the market, it continues to remain a fertile ground for academic research. Unanswered questions persist, and new challenges arise as researchers push the boundaries of quantum communication technology. New distance records are established regularly, either by advancing the capabilities of existing protocols or by introducing entirely new protocols, as seen in the case of twin-field QKD.

Furthermore, as researchers strive to achieve better performance and security, QKD not only offers a commercial reality but also serves as a vital alternative to algorithmic cryptography for achieving physical security in our increasingly interconnected and digitized world. The ongoing pursuit of better technologies in this field reflects the dedication of scientists worldwide to address the critical issues of secure communication and data protection.

## Acknowledgements

N.G. acknowledge support by the Swiss NCCR-swissMap

D.R. thanks the Galician Regional Government (consolidation of Research Units: AtlantTIC), MICIN with funding from the European Union NextGenerationEU (PRTR-C17.I1) and the Galician Regional Government with own funding through the "Planes Complementarios de I+D+I con las Comunidades Autónomas" in Quantum Communication and The European Union's Horizon Europe Framework Programme under the project "Quantum Security Networks Partnership" (QSNP, grant agreement No 101114043).